# Graphene Nanoribbon Field-Effect Transistors on Wafer-Scale Epitaxial Graphene on SiC substrates

Wan Sik Hwang[1,5,*], Pei Zhao[1], Kristof Tahy[1], Luke O. Nyakiti[3], Virginia D. Wheeler[3], Rachael. L. Myers-Ward[3], Charles R. Eddy Jr.[3], D. Kurt Gaskill[3], Joshua A. Robinson[4], Wilfried Haensch[2], Huili (Grace) Xing[1], Alan Seabaugh[1], and Debdeep Jena[1,**].

[1]Department of Electrical Engineering, University of Notre Dame, Notre Dame, IN 46556, USA
[2]IBM T. J. Watson Research Center, Yorktown Heights, NY 10598, USA
[3]U. S. Naval Research Laboratory, Washington DC, 20375, USA
[4]Materials Science and Engineering & Center of 2D & Layered Materials, Pennsylvania State University, University Park, PA 16802, USA
[5]Department of Materials Engineering (MRI), Korea Aerospace University, Gyeonggi-do, 412791, Korea

Email: **djena@nd.edu & *whwang@kau.ac.kr

**Abstract**

We report the realization of top-gated graphene nanoribbon field effect transistors (GNRFETs) of ~10 nm width on large-area epitaxial graphene exhibiting the opening of a band gap of ~0.14 eV. Contrary to prior observations of disordered transport and severe edge-roughness effects of GNRs, the experimental results presented here clearly show that the transport mechanism in carefully fabricated GNRFETS is conventional band-transport at room temperature, and inter-band tunneling at low temperature. The entire space of temperature, size, and geometry dependent transport properties and electrostatics of the GNRFETs are explained by a conventional thermionic emission and tunneling current model. Our combined experimental and modeling work proves that carefully fabricated narrow GNRs behave as conventional semiconductors, and remain potential candidates for electronic switching devices.

Implementation of 2-dimensional (2D) graphene for digital logic devices has proven challenging because of the material's zero band gap [1]. Various alternate digital logic device structures have been proposed that take advantage of interlayer tunneling, graphene-3D semiconductor heterostructure, and properties that exploit the light-like energy dispersion of carriers in 2D graphene [2-6]. From the point of view of realizing conventional field-effect transistors, well-controlled graphene nanoribbons (GNRs) mimic the excellent electrostatic properties of carbon nanotubes (CNTs) and offer hope for graphene-based digital logic devices [7, 8]. The ultrathin body can enable scaling down to 10 nm or below while still keeping short-channel degradation effects at bay. GNRs suffer from edge-roughness scattering effects compared to CNTs, but GNRs provide better large-area scalability, planar fabrication opportunity, and heat dissipation capacity than CNTs [9]. The availability of broken bonds at the edges provides a window of opportunity for chemical doping [10], which remains difficult in CNTs due to saturated $sp^2$ chemical bonds. A number of "beyond-CMOS" devices, such as the GNR tunneling field-effect transistor (TFET) [11] can be realized if controlled GNRs can be fabricated on large-area substrates. Thus, progress in the fabrication and characterization of wafer-scale GNRs stands to potentially enable a host of applications in the future.

The creation of controlled band gaps by quantum confinement of carriers in GNRs remains a significant challenge [12~21]. To date, graphene nanoribbon field effect transistors (GNRFETs) down to 10 ~ 20 nm channel width have been fabricated from exfoliated graphene [13, 14] and chemical vapor deposition (CVD) grown graphene [15, 16] using conventional top-down lithography and etching methods. Bottom-up techniques such as chemically derived

GNRFETs down to sub-5 nm width have been fabricated, and show substantial band gaps with $I_{ON}/I_{OFF}$ ~$10^6$ at room temperature [17]. GNRFETs have also been fabricated by unzipping CNTs [18-20]. More recently, GNRs down to 5 nm has been directly grown on SiC substrates using ion implantation followed by laser annealing [21]. But the bottom-up techniques are not yet site-controlled and reproducible, and are currently incompatible with conventional lithographic processes for circuit implementations.

Epitaxial graphene (EG) grown on single-crystal, semi-insulating SiC wafers satisfy many of the above criteria [22, 23]. Furthermore, devices based on EG require fewer processing steps and are more immune to contamination compared to CVD-grown large-area graphene due to the absence of a transfer process. GNRFETs can mimic properties of CNTFETs and remove needs of alignment and random mixtures of metallic and semiconducting channels. The major challenge in realizing GNRs is in achieving ~5 nm widths with smooth edges. In this pursuit, GNRFETs stand to benefit from recent process developments in Silicon FinFET technology, in which arrays of ~5 nm wide Si fins have been demonstrated with robust structural integrity [24]. Process variation challenges of such narrow fins have been addressed for next-generation CMOS technology [25].

Despite the importance of EG, substantial energy gaps have not yet been demonstrated in GNRFETs made in EG on SiC [26]. Furthermore, there are no studies that correlate experimentally measured transport properties and theoretical models for EG-GNRs. In this work, we report the fabrication of top-gated ~10 nm wide GNRFETs by lithography on large area EG on SiC substrates. We observe for the first time, the opening of a substantial energy gap inversely proportional to the GNRFET width of EG-GNRs. By relating the measured transport with theoretical modeling, we find that the transport properties of narrow epi-GNRs are similar to well-behaved narrow-bandgap semiconductors, contrary to carrier localization effects reported extensively in wider GNRs fabricated on exfoliated graphene [27-31]. The reasons for these observations will be discussed.

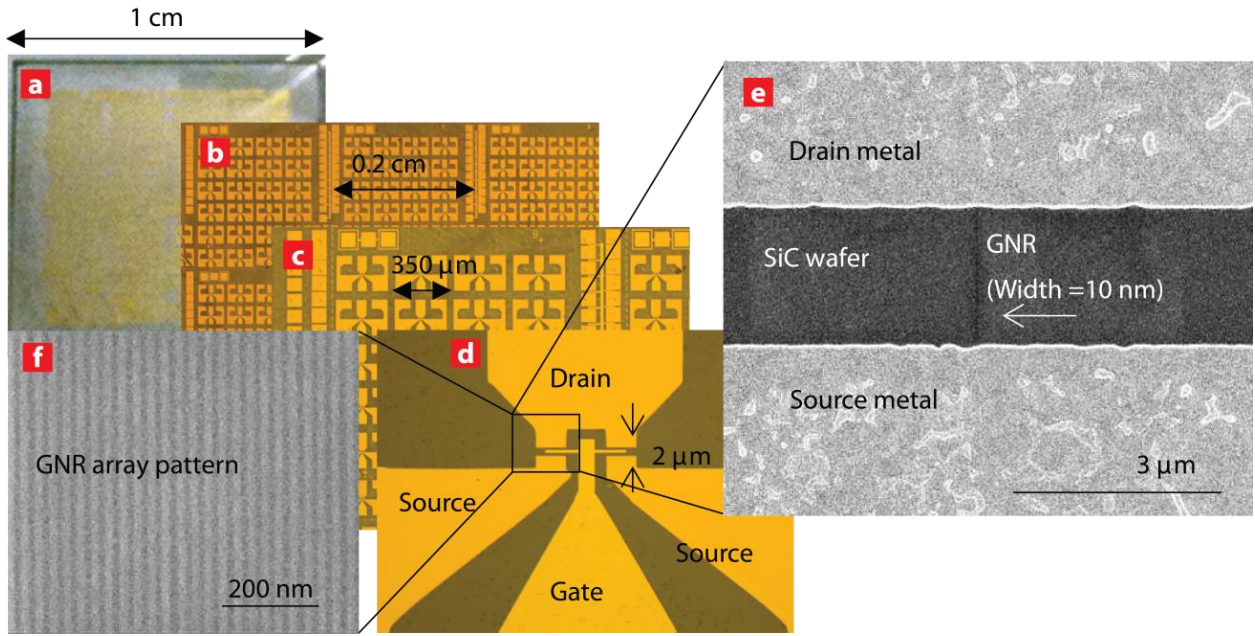

**Fig. 1.** (a) ~ (d) Optical microscope image of epitaxial graphene nano-ribbon (GNR) FETs on wafer size SiC substrate. (e) Scanning electron microscope (SEM) image of GNR having 10 nm widths with source and drain metal. (f) SEM image of HSQ array ribbon patterns, consisting of 13 nm line with and 17 nm space, showing no deformation and collapse. The HSQ patterns play a role as a mask to etch graphene during $O_2$ Plasma. Finally, GNR remains after removing the HSQ mask.

The starting material in this work was epitaxial-graphene grown on a 4 inch diameter Si-face 6H-SiC substrate. The epitaxial growth conditions are described in earlier reports [22] and this epitaxial graphene on SiC is expected to have lower residual charge than transferred graphene ($2\sim5 \times 10^{11}$ cm$^{-2}$) on $SiO_2$ due to the absence of transfer process [32, 33]. Figure 1 shows the final device images including single GNR and arrays of GNRs. Hydrogen silsesquioxane (HSQ), a negative-tone electron-beam resist, was used to fabricate GNRs of varying widths, down to ~10 nm. The gate stack consists of 15 nm HSQ followed by atomic layer deposited (ALD) 30 nm $Al_2O_3$ at 200 $^oC$ and Cr (5 nm) / Au (100 nm) using electron-beam evaporator. The GNRs are connected to the two-dimensional (2D) graphene area and the source /drain contact metal of Cr (5nm) / Au (100nm) sits on the 2D area forming ohmic contact with zero band gap, so the energy barrier for electrons entering the GNR is half the GNR band gap. Form the transfer characteristics, extrinsic field-effect mobility was extracted as 800~1000 $cm^2$/V.s at maximum transconductance. The contact resistance was not accounted for in the mobility extraction and the contact resistance extracted from TLM patterns is around $10^4$ Ω.µm. Details of the HSQ process and device processing flow have been discussed earlier [34].

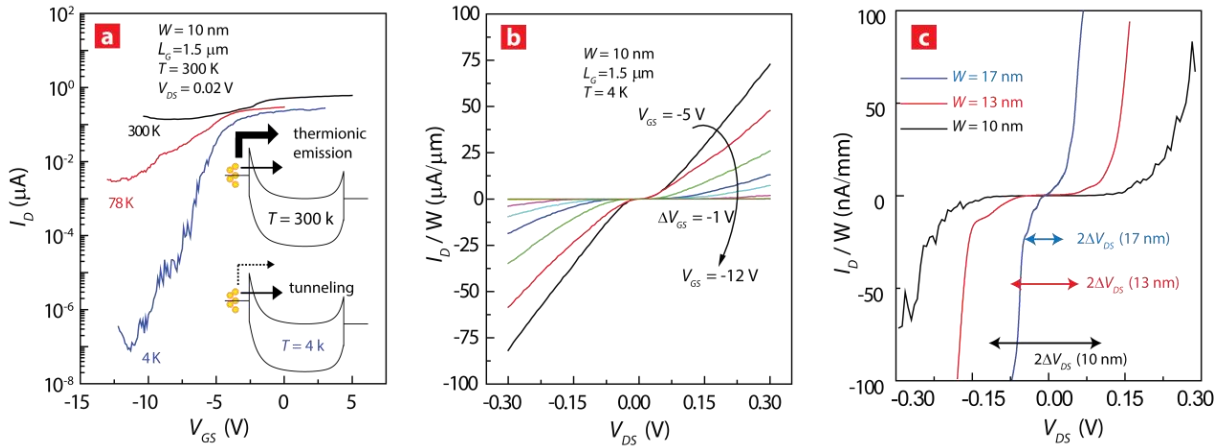

**Fig. 2.** (a) $I_D$ *versus* $V_{GS}$ of 10 nm width GNRFETs at various temperatures. $I_D$ is dominated by thermionic emission at 300 K, while it is controlled by band to band tunneling at 4 K since thermionic emission is suppressed. (b) Family $I_D$ *versus* $V_{DS}$ of 10 nm GNRFETs at various $V_{GS}$ at 4 K showing clearly on/off state. (c) $I_D$ *versus* $V_{DS}$ depending on different GNR width with $V_{GS}$ fixed in the charge neutral voltage at 4 K.

Figure 2(a) shows the measured drain current $I_D$ as a function of the gate bias $V_{GS}$ of 10 nm GNRFETs at three different temperatures. The gate modulation (ratio of $I_{ON}/I_{OFF}$) of the drain current is about 10 X. The relatively high $I_{OFF}$ observed at 300 K is due to thermionic emission current from the source contact. For a 10 nm wide GNR, the energy gap is $E_g \sim 0.14$ eV, which leads to a Schottky barrier height of $qf_B \sim E_g/2 \sim 70$ meV. This is only slightly smaller than $\sim 3kT$ at room temperature, implying a large thermionic emission current over the barrier since $I_{off} \sim \exp[-qf_B/kT]$. This temperature dependence is accentuated at lower temperatures, because $I_{on}$ stays relatively constant whereas $I_{off}$ reduces by several orders of magnitude due to the reduction of the thermionic emission current. This results in an increase of $I_{on}/I_{off} \rightarrow 10^6$ at 4 K as shown in Fig 2(a). At this low temperature, the Fermi-Dirac tail of the electron distribution in the source is severely curtailed, and electrons have to tunnel through the energy gap of the GNR. This band-to-band transport mainly happens across the barrier formed at the contact, not entire length of the device.

This strong temperature dependence of $I_D$ in the GNR FET shown in Fig. 2(a) is distinctly different from 2D FETs [34], revealing a presence of energy gap. The family $I_D$ versus $V_{DS}$ of 10 nm GNRFETs in Fig. 2(b) clearly shows the "turn-on" and "turn-off" region depending on location of Fermi level, which is tuned by the gate bias, $V_{GS}$. Figure 2(c) shows $I_D$ vs. $V_{DS}$ for GNRs of three different widths (10, 13, and 17 nm) at a $V_{GS}$ biased near the charge neutral voltage. The current-voltage curve in Fig. 2(c) is characteristic of back-to-back metal-semiconductor Schottky diodes, and the turn-on window is a measure of the Schottky barrier height. It is observed that as the widths of GNRs decrease the size of the low-conductance window increases. The energy gap is inversely proportional to the widths of GNR FET channels.

In order to measure the band gap quantitatively, a more comprehensive approach entails measuring the conductance map as a function of $V_{DS}$ and $V_{GS}$. The results of such measurements are discussed next.

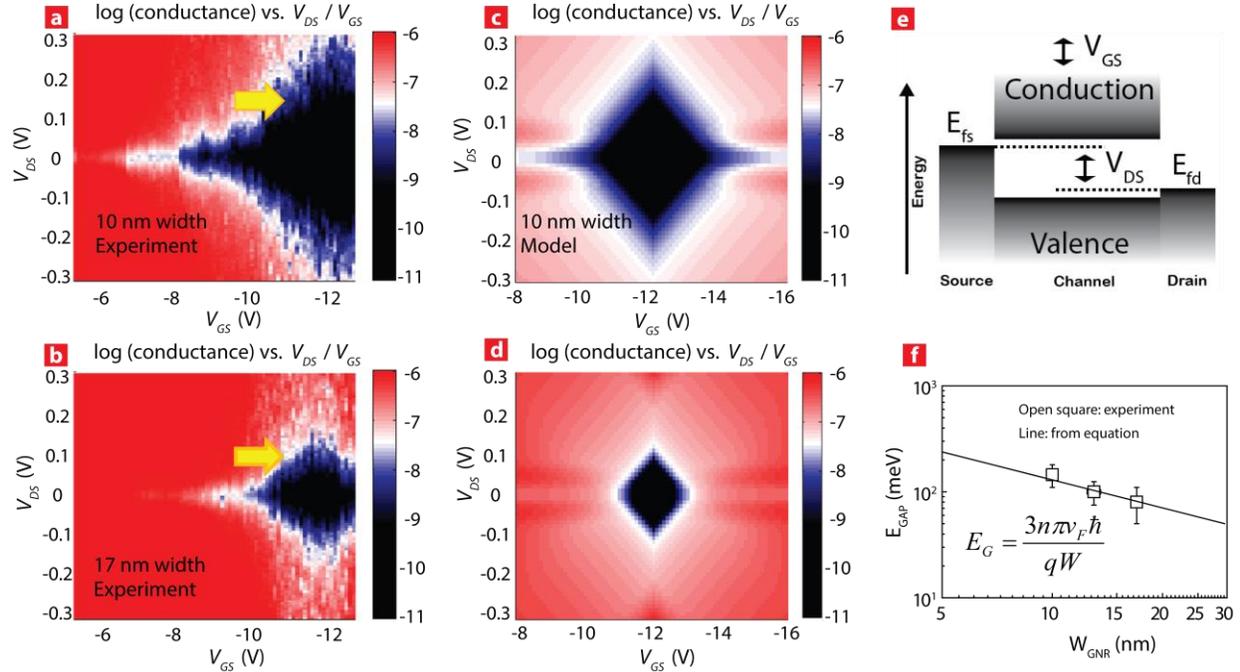

**Fig. 3.** The differential conductance of two representative GNRFETs of 10 nm width (a) and 17 nm width (b) as a function of $V_{DS}$ and $V_{GS}$ at 4 K. Modeling results of two different widths of 10 nm (c) and 17 nm (d) GNRFETs which is corresponding (a) and (b) respectively. The black (dark) color represents a low conductance as indicated by color map. (e) The energy band diagram was used in the model which was developed based on Schottky barrier. (f) Extracted band gap of GNRFET vs. width of GNR. The linear line was predicted by the model. The deviation of this GNRs width is around 0.5 nm by SEM.

Figure 3 shows the 4 K conductance versus $V_{DS}$ and $V_{GS}$ for two representative GNRFETs of 10 nm (a) and 17 nm (b). The conductance is shown as a color in a logarithmic scale with red representing high (on-state) and black as low conductance (off-state). For a fixed drain bias, scanning the gate voltage results in a sharp transition from conducting to insulating states. For example, the 4 K scan in Fig. 3(a) is for a drain bias of 20 mV in which the transition is seen in the region -8 V < $V_{GS}$ < -6 V. Similarly, for a fixed gate bias, scanning the drain voltage reveals the back-to-back Schottky behavior shown in Fig. 3(a) and (b). Realizing that the channel is no different from a traditional semiconductor (albeit with a small band gap), an accurate method to extract the energy band gap is to model the *entire* conductance map using traditional semiconductor transport equations, accounting for *both* thermionic emission and tunneling current components. Because the contacts are Schottky barriers of height half the energy band gap, and tunneling depends on the band gap, modeling the dependence of the conductance maps on the GNR widths enables an accurate extraction of the energy band gap.

The details of the hybrid thermionic emission / tunneling transport model are provided in the supplementary materials accompanying this paper. The modeled conductance versus $V_{DS}$ / $V_{GS}$ for GNRs of widths 10 nm and 17 nm are shown in Fig. 3(c) and (d) alongside the measured experimental results. The simple textbook-model of transport captures the entire shape of the conductance map. Note that no localization or quantum-dot type hopping transport was used in the model. The band-edge fluctuations that may result from the line-edge roughness of the GNRs cause a smearing of the on-off state transition observed in the experimental data compared to the sharp transitions predicted by the model. Such fluctuations are a measure of the disorder in the GNR, but they are minimal compared to the overall characteristics, which are captured from a band-transport picture. We note that such fluctuations are not limited to GNRs alone; indeed, they can be observed in most narrow-gap semiconductor FETs. We do not observe any Coulomb diamonds and charging effects as reported earlier [27 ~ 30]. The reason for this difference from earlier reports is twofold. First, because the GNRs reported here are among the *narrowest* reported to-date fabricated by lithographic techniques, the corresponding energy gaps are among the highest. Second, the epitaxial graphene on SiC substrates reveal smaller potential variation of 12 meV [35] than that of transferred graphene on $SiO_2$ of 59~77 meV [27~33]. The potential fluctuations due to charged impurities can localize carriers if the band gap is small, whereas a larger band gap coupled with low residual impurity density enables conventional band-edge transport

.

The dependence of the band gaps on the GNR widths extracted from Fig. 3(a) and 3(b) is shown in Fig. 3(f) and is in good agreement with a conventional model of GNR band gaps [36]. An energy band gap of $E_g$~0.14 eV was achieved by scaling to a GNR width of ~10 nm. Based on this observation, an $E_g$~0.3 meV could be achieved by narrowing the GNR width down to 5 nm, a distinct possibility in the future based on FinFET technology [24]. Since the transport model is based on thermionic emission and tunneling, it may be used for predicting the behavior of GNR FETs of different channel lengths. This is a crucial test for a transport model: tunneling from the source to drain is heavily dependent on the S/D separation at small voltages, whereas thermionic emission is not. Using the model and the corresponding experimental measurement, we can verify the accuracy of the model further. To do so, we performed transport measurements on GNRFETs with varying S/D distances and compared with the predictions from the model, as discussed next.

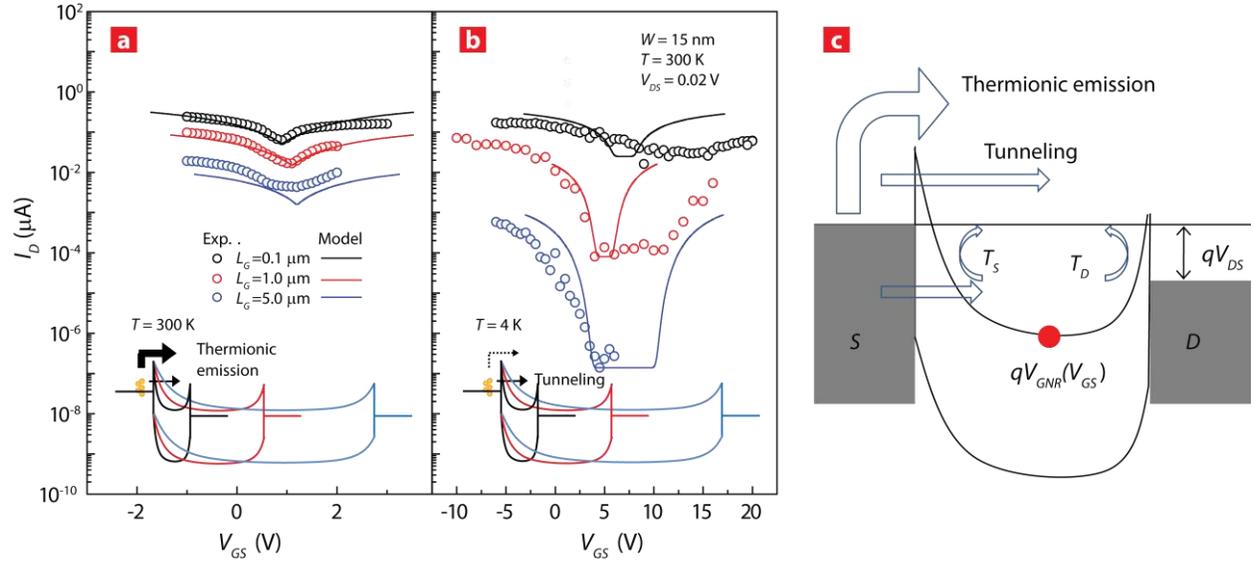

**Fig. 4.** $I_D$ *versus* $V_{GS}$ of 15 nm width GNRFETs depending on different gate length at 300 K (a) and at 4 K (b). The inset (right bottom) of each figure shows a schematic image of major transport mechanism depending on temperature. The $I_D$ dependence on channel length can be explained by the channel resistance at 300 K (a) and band to band tunneling at 4 K (b). (c) General concept of current mechanism which was taken account in this modeling at both (a) and (b).

Figures 4(a) & (b) show the $I_D$ *vs* $V_{GS}$ of 15 nm wide GNRFETs for three gate lengths: 5 μm, 1 μm, and 0.1 μm measured at 300 K (a) and 4 K (b). It is observed that $I_D$ increases as the gate length decreases at both 300 K and 4 K. As the gate length decreases the gate modulation remains relatively constant at 300 K in Fig. 4(a), whereas the gate modulation changes exponentially at 4 K in Fig. 4(b) since the conduction is dominated by tunneling." The corresponding predictions based on the hybrid thermionic emission / tunneling current model are shown as solid lines in Fig. 4(a) and (b). The energy band diagram corresponding to the model is shown in Fig 4(c). The model can capture most of the experimentally observed behavior, further

lending credence to the claim that transport in the GNRs is band-like, and hopping and localization effects need not be invoked to explain the device behavior.

Several recent studies have associated charge transport in GNRs with hopping conductivity and quantum dot behavior [27 ~ 31], and not by conventional conduction mechanisms. We discuss these earlier observations in light of our observations. The observations can be resolved by paying careful attention to GNR widths, surface potential variation of graphene, GNR edge roughness, and device operation regimes. First, in earlier reports, GNR widths range from 30-100 nm, which will lead to energy band gaps less than ~50 meV. This energy gap is comparable to the electron-hole puddle surface potential variations, which have been reported to be around 50 ~ 80 meV [27 ~ 31] for the graphene/$SiO_2$ interface, but only 12 meV for graphene/SiC interface [35]. When the disorder potential variations are of the order of, or more than the energy band gap, it constitutes a severe perturbation of transport properties. Furthermore, the high density of such fluctuations in graphene/$SiO_2$ interfaces exacerbates the localization of carriers leading to hopping transport. On the contrary, when the energy gap is larger, as obtained with narrower GNRs, the potential disorder behaves like a weak fluctuation, similar to ionized impurity doping in traditional semiconductors. The residual charge densities in GNRFETs in previous reports [27 ~ 31] is expected to be high, since the GNRs were fabricated from exfoliated graphene transferred on to $SiO_2$ substrates. In addition, the HSQ mask (15 nm height) produced by EBL to etch graphene results in smooth epi-graphene GNRs, whose edge roughness is estimated to be less than ~0.35 nm through root mean square (RMS) estimation of the width by image processing [15] while one of edge roughness from previous work is around 4 nm [28]. Finally, the device operation regime in the conductance map reported in this work spans hundreds of meV range, unlike ~50 meV ranges reported earlier.

The experimental results reported here and the above discussions suggest that even though there is potential fluctuation caused by either line edge roughness or potential inhomogeneity, the behavior of epi-GNR FETs is indeed *no different* from any conventional narrow-bandgap semiconductor. Most effects like ratio of $I_{ON}/I_{OFF}$ observed in the transport of previously reported GNRs mimic those of disordered or heavily doped narrow-bandgap semiconductors [37], and as GNRs become narrower and cleaner, their intrinsic properties and electrostatic advantages will make them highly attractive for electronic devices in the future.

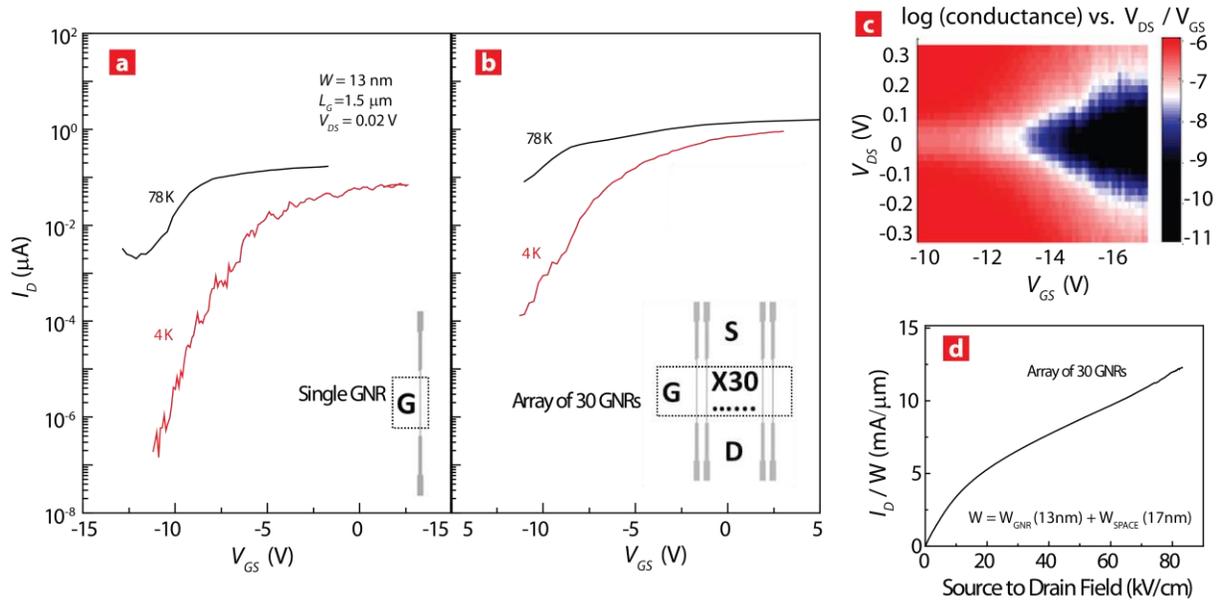

**Fig. 5.** $I_D$ versus $V_{GS}$ of single GNRFET (A) and array GNRFET (B) at both 4K and 78K. (C) The differential conductance of an array of 30 GNRFETs having 13 nm widths. (D) Maximum drain current density of the array of 30 GNRFETs of 12 mA/μm (scaling by the total channel width: 30 x 13 nm line width + 30 x 17 nm space = 900 nm. A maximum drain current density of 28 mA/μm can be achieved if the current density is divided by only active GNR area (30 x 13 nm = 390 nm). The drain current $I_D$ of ~$10^{-12}$ A current was obtained after etching GNRs showing the current conduction is indeed through the GNRs and not the substrate.

Array GNRFETs consisting of parallel arrays of 30 GNRs were fabricated and the $I_D$ versus $V_{GS}$ of the 30-array GNRFET was compared with that of single GNRFET, all with the same 13 nm width as shown in Fig. 5(a) and (b). The inset of Fig. 5(b) shows the schematic image of an array GNRFET, comparing the inset of a single GNRFET in Fig. 5(a). It shows that the individual device performance of array GNRFET is *preserved in the array structure*. The increase of drain current is one of the benefits of array GNRFETs. We also observe a high maximum drain current density of ~12 mA/μm considering the total channel width. Such high current drives have never been reported in *any* semiconductor device [38]. If we consider only the active ribbon width, the maximum high drain current density becomes 28 mA/μm, a value that may be approached by changing the pitch of the GNR array. We attribute this high current carrying capability to the high electrical and thermal conductivity of the GNR channels due to the absence of lateral scattering, coupled with the excellent thermal conductivity of the underlying SiC substrate. The high current drives are attractive from many viewpoints: for high-performance transistors with fast switching, and possibly for integrated interconnects.

In summary, we report results of the first top-gated 10 nm width GNRFETs on large-area epitaxial graphene exhibiting exceptionally high drive currents, the opening of a substantial band gap, and an increase of drain current by exploiting FET arrays. The narrow GNR width in the range of 10 nm and the epitaxial platform enables conventional current flow mechanism without introducing hopping effect and quantum dot behavior. The measured transport dependence over the entire parameter space (GNR width, gate length, temperature) is explained accurately by invoking a *single* conventional thermionic emission + tunneling model. With further scaling of the widths of wafer-scale clean GNRFETs, graphene based transistors can show promising potential for logic applications.

This work was supported by the Semiconductor Research Corporation (SRC), Nanoelectronics Research Initiative (NRI) and the National Institute of Standards and Technology (NIST) through the Midwest Institute for Nanoelectronics Discovery (MIND), STARnet, an SRC program sponsored by MARCO and DARPA, and by the Office of Naval Research (ONR) and the National Science Foundation (NSF). LON is grateful for postdoctoral support from the ASEE.

[34] W. S. Hwang, K. Tahy, L. O. Nyakiti, V. D. Wheeler, R. L. Myers-Ward, C. R. Eddy Jr., D. K. Gaskill, H. Xing, A. Seabaugh, and D. Jena, *Fabrication of top-gated epitaxial graphene nanoribbons FETs using hydrogen-silsesquioxane*, J. Vac. Sci. Technol. B **30**, 03D104 (2012).

[35] A. E. Curing, M. S. Fuhrer, J. L. Tedesco, R. L. Myers-Ward, C. R. Eddy Jr., D. K. Gaskill, *Kelvin probe microscopy and electronic transport in graphene on SiC (0001) in the minimum conductivity regime,* Appl. Phys. Lett. **98**, 243111 (2011).

[36] B. Trauzettel, D. V. Bulaev, D. Loss, and G. Burkard, *Spin qubits in graphene quantum dots*, Nature **3**, 192-196 (2007).

[37] H. A. Nilsson, P. Caroff, C. Thelander, E. Lind, O. Karlstrom, and L. –E. Wernersson, *Temperature dependent properties of InSb and InAs nanowire field-effect transistors*, Appl. Phys. Lett. **96**, 153505 (2010).

[38] K. Shinohara, D. Regan, A. Corrion, D. Brown, Y. Tang, J. Wong, G. Candia, A. Schmitz, H. Fung, S. Kim, and M. Micovic, *Self-aligned-gate GaN-HEMTs with heavily-doped $n^+$-GaN ohmic contacts to 2DEG*, IEEE International Electron Devices Meeting, 27.2.2-27.2.4 (2012).
19

**Supplementary Materials supporting the main manuscript**

# Graphene Nanoribbon Field-Effect Transistors on Wafer-Scale Epitaxial Graphene on SiC substrates

Wan Sik Hwang[1,5,*], Pei Zhao[1], Kristof Tahy[1], Luke O. Nyakiti[3], Virginia D. Wheeler[3], Rachael. L. Myers-Ward[3], Charles R. Eddy Jr.[3], D. Kurt Gaskill[3], Joshua A. Robinson[4], Wilfried Haensch[2], Huili (Grace) Xing[1], Alan Seabaugh[1], and Debdeep Jena[1,**].

[1]Department of Electrical Engineering, University of Notre Dame, Notre Dame, IN 46556, USA
[2]IBM T. J. Watson Research Center, Yorktown Heights, NY 10598, USA
[3]U. S. Naval Research Laboratory, Washington DC, 20375, USA
[4]Materials Science and Engineering & Center of 2D & Layered Materials, Pennsylvania State University, University Park, PA 16802, USA
[5]Department of Materials Engineering (MRI), Korea Aerospace University, Gyeonggi-do, 412791, Korea

Email: **djena@nd.edu & *whwang@kau.ac.kr

## 1. Analytical Model for GNRFETs simulation

The top-gated transistor structure has been carefully studied before [1]. The spatial variation of the conduction band edge profile from the source to drain (the coordinate *x*) can be described by the analytical closed form expressions

$$E_{C,left} = \phi_B - (\phi_B - qV_{GNR})\frac{2}{\pi}\arccos(\exp(\frac{-x\pi}{2t_{ox}})) \tag{1}$$

$$E_{C,right} = (\phi_B - qV_{DS}) - (\phi_B - qV_{GNR} - qV_{DS})\frac{2}{\pi}\arccos(\exp(\frac{(x-L)\pi}{2t_{ox}})) \tag{2}$$

where $t_{ox}$ is oxide thickness, $V_{DS}$ is the applied drain voltage, and $L$ is the channel length. The two expressions capture the energy band bending on the left (source-side) and the right (drain-side). The Schottky Barrier height $\Phi_B$ is assumed to be half of the bandgap $\Phi_B=E_G/2$, and the unintentional doping in channel is captured by $V_{G,min}$. The local potential of the GNR channel $qV_{GNR}$ can be derived from the electrostatics equation [2] as a function of the applied bias:

$$qV_{GNR} = \frac{C_{ox}(V_G - V_{G,\min}) + C_{GD}V_D + C_{GS}V_S + Q_{ch}}{C_{ox} + C_{GS} + C_{GD}} \quad (3)$$

where $qV_{GNR}$ is local potential of the GNR channel, $C_{OX}$ is gate capacitance, $V_G$ is the gate voltages, $V_{G,min}$ is shift of minimal conduction voltage caused by unintentional doping in the GNR, $C_{GD}$ is gate-drain capacitance, $V_D$ is the drain voltages, $C_{GS}$ is gate-source capacitance, $V_S$ is source voltage, $Q_{ch}$ is carrier charge density.

The carrier charge density is calculated by integrating over the density of states:

$$Q_{ch} = q\int \frac{1}{2}\left[ D(E - qV_{GNR} - \frac{E_G}{2})f(E - E_F) + D(E - qV_{GNR} - \frac{E_G}{2})f(E - E_F - qV_D)\right]dE \quad (4)$$

Where $D(E)$ is the density-of-states of GNR, and $f(x)=1/(1+\exp[x])$ is the Fermi-Dirac function. The effect of parasitic capacitances can be included with the gate-source and gate-drain overlap capacitances $C_{GS}$ and $C_{GD}$. In the GNRFETs discussed in this work, the gate length and S/D distance are several micrometers long. Thus the effect of the parasitic capacitances is negligible. Using equations (1) and (2), we obtain $qV_{GNR}$ as a function of $V_G$ and $V_{DS}$. The band profile is then used as the input parameter for the calculation of the drain current.

The current is calculated by summing the current spectrum over the entire energy window:

$$I = \frac{2q}{h} \int Tr(E)\{f[E - E_{FS}] - f[E - E_{FD}]\}dE \tag{5}$$

where the total current includes the thermal emission current over the barrier, and the tunneling current thought the bandgap. *Tr(E)* is the overall energy dependent transmission coefficient from the source to drain. For thermal emission, we assume that *Tr(E)*~1 without any quantum reflection. The transmission coefficient *Tr(E)* due to tunneling is

$$Tr(E) = \frac{T_S(E)T_D(E)}{T_S(E) + T_D(E) - T_S(E)T_D(E)}, \tag{6}$$

where $T_S(E)$ and $T_D(E)$ are two coefficients decided by the source and drain barriers separately. They are calculated using the WKB approximation:

$$T_{S(D)}(E) = \exp\left(-2\int_{xinit}^{xfinal} k_x(x)dx\right), \tag{7}$$

where

$k_x = \sqrt{\frac{(E - E_C(x) - E_G/2)^2}{(v_F \hbar)^2} - k_n^2}$ is the position and energy dependent momentum,

$k_n = 3n\pi/W$ is the quantized transverse momentum, and the bandgap is $E_G = 3n\pi v_F \hbar / qW$.

The hole current is calculated using the same formalism by substituting the conduction band edge $E_C(x)$ with the valence band edge $E_V(x) = E_C(x) - E_G$.

The above equations assume ballistic transport. In a long channel transistor, the effects of phonon scattering, impurity scattering, and edge roughness need to be considered. The optical phonon scattering can be ignored here since $V_{DS}$ is only 20 mV, which is much less than the 160meV phonon energy. Other scattering mechanisms are represented by a phenomenological parameter mean free path $\lambda$, and the total reduction of current is $\lambda / (\lambda+L)$ [2]. Edge roughness is believed to have a strong effect on transport with additional gap states induced into the bandgap region. In the analytical model, we assume that the edge roughness produces a band gap tail with a band gap reduction around 10% [3].

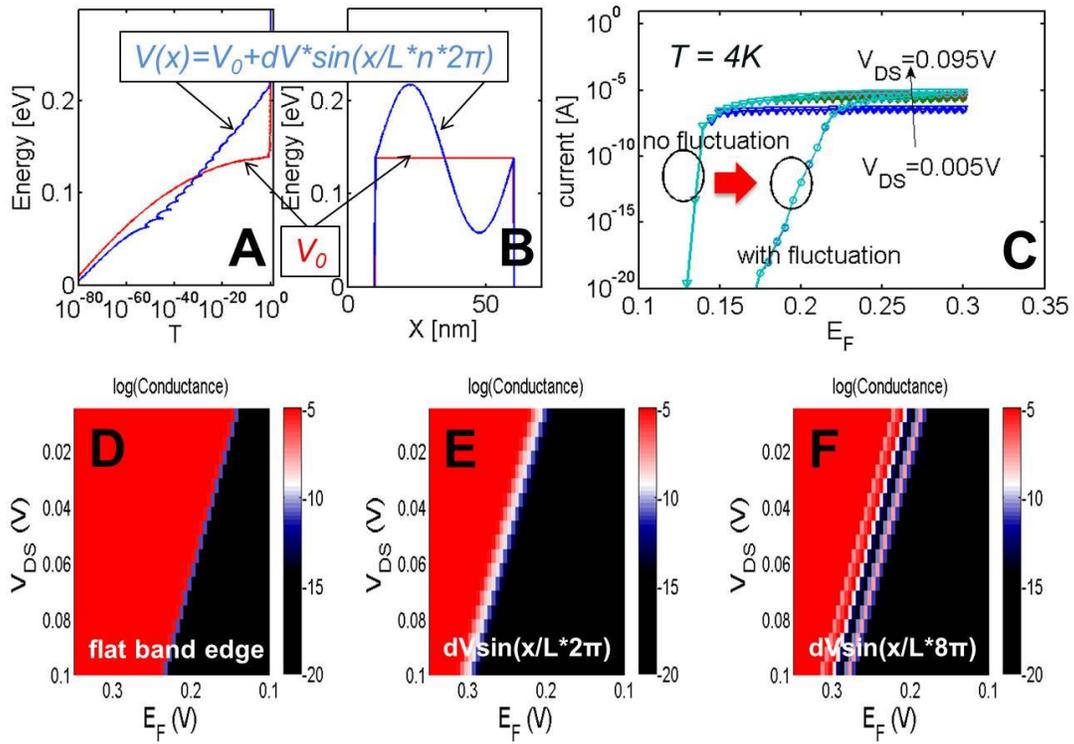

**Fig. S1.** Energy band edge profile to mimic the fluctuation in Fig. 3 mapping. The quantum interference due to potential fluctuation can be seen in (A) and (B). The tunneling current in (C) shows that with fluctuation the sharpness of "turn-off" is degraded. (D), (E) and (F) plot the conductance map with different fluctuation function. The 'diamond' edge expands and smears. The well spaced traps (sinusoid fluctuation) leads to oscillatory conductance as expected.

As shown in Figure 4 in the main text, the analytical model captures the 'diamond' shape in the conductance map. The single model predicts the dependence of the bandgap opening at different

GNR widths accurately. In the real device, the edge of the "diamond" conductance map is not sharp. The reason is attributed to a disorder potential in the channel. A simple model for the disorder potential is used to understand its effect on the conductance spectra. The non-uniform tunneling barrier effect is shown in Fig. S1.

The non-ideal potential profile modeled as

$$V(x) = V_O + dV \sin(\frac{x \times n \times 2\pi}{L})$$

Where a mimic fluctuation $dV \sin(\frac{x \times n \times 2\pi}{L})$ is added with the ideal potential $V_0$. The quantum interference due to the potential fluctuation can be seen in Fig. S1(A) & (B), and the corresponding tunneling current is shown in Fig. S1(C). As the fluctuation increases, the degradation of the "turn-off" sharpness and smearing become obvious as shown in Fig. S1 (D) ~ (F). Thus the device model is also capable of incorporating the effect of edge-roughness on the transistor characteristics. This model can be developed further for deterministic device design purposes, and simultaneously offer insights to the nature of the GNRs.

2. **TEM analysis of array GNRs**

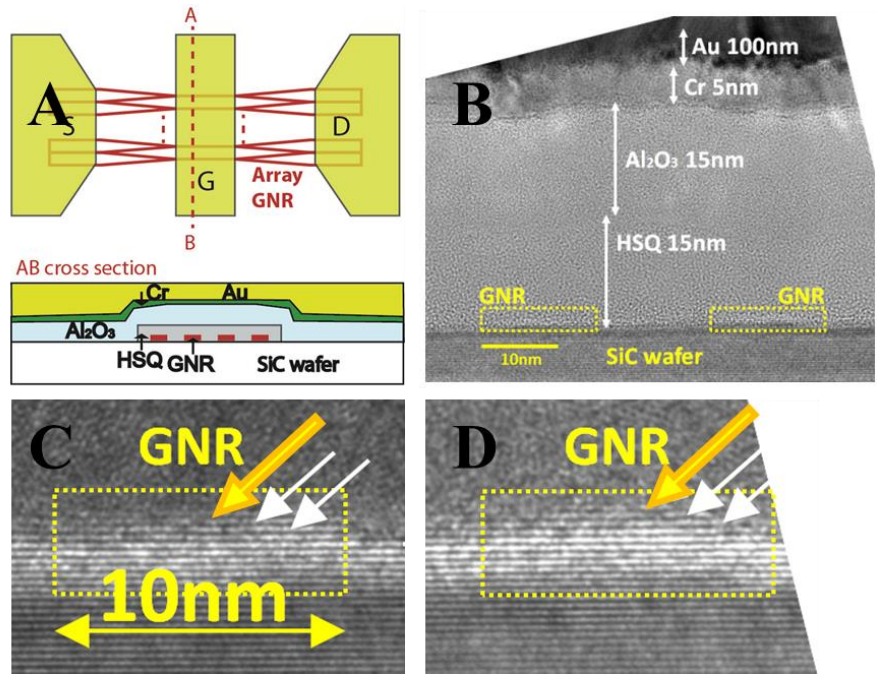

**Fig. S2.** (A) Schematic image of TEM taken area, (B) TEM image of gate stack consisting of Au/Cr/Al$_2$O$_3$/HSQ/GNR on SiC wafer. (C) & (D) Focused TEM image of GNR. The white arrow indicates the isolated carbon layer different from SiC wafer and the top layer is monolayer graphene since all of the GNR were formed at the terraces not at the step on SiC wafer.

To understand the microscopic nature of the GNRs, TEM images were taken in the regions of GNR array structures. They are shown in Fig. S2. The lattice image of the SiC substrate is clearly visible, and the GNRs could be identified in the array regions because of their periodicity. No abnormal interfacial layers are observed between the HSQ and the GNRs in the TEM image, supporting the stable device behavior.

[1] Jimenez D. A current–voltage model for Schottky-barrier graphene based transistors. *Nanotechnology* **19**, 345204 (2008)